\documentclass[sigconf]{acmart}

\usepackage{adjustbox}
\usepackage{multirow}
\newcommand{\etal}{\textit{et al.}}
\usepackage{balance}
\usepackage[outdir=./]{epstopdf}

\AtBeginDocument{
 }

\copyrightyear{2024}
\acmYear{2024}
\setcopyright{rightsretained}
\acmConference[SIGIR '24]{Proceedings of the 47th International ACM SIGIR Conference on Research and Development in Information Retrieval}{July 14--18, 2024}{Washington, DC, USA}
\acmBooktitle{Proceedings of the 47th International ACM SIGIR Conference on Research and Development in Information Retrieval (SIGIR '24), July 14--18, 2024, Washington, DC, USA}
\acmDOI{10.1145/3626772.3657976}
\acmISBN{979-8-4007-0431-4/24/07}

\begin{document}

\title[Distance Sampling-based Paraphraser Leveraging ChatGPT for Text Data Manipulation]{Distance Sampling-based Paraphraser \\Leveraging ChatGPT for Text Data Manipulation}

\author{Yoori Oh}
\authornote{Both authors contributed equally to this research.}
\authornotemark[0]
\affiliation{%
  \institution{Department of Intelligence and Information}
  \institution{Seoul National University}
  \state{Seoul}
  \country{Republic of Korea}
}
\email{yoori0203@snu.ac.kr}

\author{Yoseob Han}
\authornotemark[1]
\affiliation{%
  \institution{Department of Electronic Engineering}
  \institution{Department of Intelligent Semiconductors}
  \institution{Soongsil University}
  \state{Seoul}
  \country{Republic of Korea}
}
\email{yoseob.han@ssu.ac.kr}

\author{Kyogu Lee}
\affiliation{%
  \institution{Department of Intelligence and Information}
  \institution{Interdisciplinary Program in Artificial Intelligence}
  \institution{AI Institute}
  \institution{Seoul National University}
  \state{Seoul}
  \country{Republic of Korea}
}
\email{kglee@snu.ac.kr}


\begin{abstract}
There has been growing interest in audio-language retrieval research, where the objective is to establish the correlation between audio and text modalities. However, most audio-text paired datasets often lack rich expression of the text data compared to the audio samples. One of the significant challenges facing audio-text datasets is the presence of similar or identical captions despite different audio samples. Therefore, under many-to-one mapping conditions, audio-text datasets lead to poor performance of retrieval tasks. 
In this paper, we propose a novel approach to tackle the data imbalance problem in audio-language retrieval task. To overcome the limitation, we introduce a method that employs a distance sampling-based paraphraser leveraging ChatGPT, utilizing distance function to generate a controllable distribution of manipulated text data. For a set of sentences with the same context, the distance is used to calculate a degree of manipulation for any two sentences, and ChatGPT's few-shot prompting is performed using a text cluster with a similar distance defined by the Jaccard similarity. Therefore, ChatGPT, when applied to few-shot prompting with text clusters, can adjust the diversity of the manipulated text based on the distance. The proposed approach is shown to significantly enhance performance in audio-text retrieval, outperforming conventional text augmentation techniques. 
\end{abstract}



\maketitle

\begin{figure}[t!]
	\centering
	\includegraphics[width=0.44\textwidth]{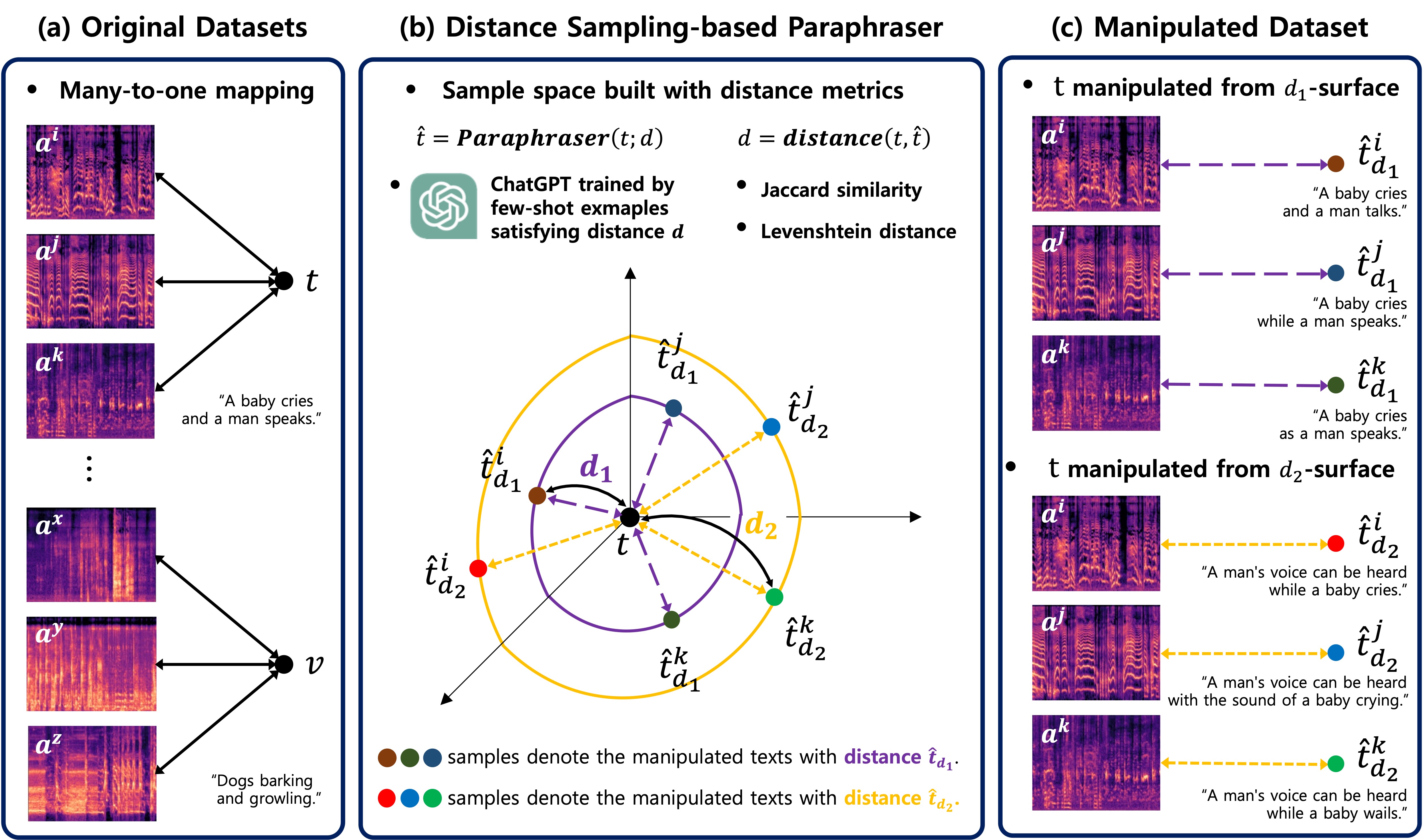}
        \caption{Overview of proposed distance sampling-based paraphraser. (a) Original audio-text dataset pairs, (b) distance sampling-based paraphraser to manipulate text data, and (c) new audio-text dataset pairs satisfying an unique mapping. }
	\label{fig:paraphraser} 
\end{figure}

\section{Introduction}
The growing interest in multimodal learning has led to remarkable advancements in vision-language representation learning. This trend has inspired an increasing number of researchers to delve into the relationship between audio and language. Fundamental tasks in this area include audio-text retrieval and audio captioning, both of which necessitate the use of audio and corresponding text-captioned data in order to perform multimodal learning. As a result, there is a pressing need to generate caption data that accurately describes audio content. The largest human-annotated dataset that can be used to address the needs is AudioCaps\cite{kim2019audiocaps}, which has become a benchmark set for evaluating downstream tasks in an audio-language domain. Nevertheless, 
the audio-language domain is still significantly inferior in terms of data size. While vision-language area benefits from access to available billion-scale datasets, the largest dataset in the audio-language area, AudioCaps\cite{kim2019audiocaps}, consists of only 46,000 samples. A simple way to increase data utilization is to collect additional audio and provide annotations, but this approach is very difficult and labor-intensive because it requires humans to listen to the audio and continuously generate captions with a variety of nuanced expressions.

There are significant challenges with labor-intensive captioning datasets, including vision or audio, collected by annotators. Figure \ref{fig:paraphraser}(a) shows a critical issue with many-to-one mapping that arises in audio-text caption datasets, and similar problems also occur in vision-language caption datasets. In the audio-language domain, many-to-one mapping samples exists, meaning that distinct audio samples are associated with the same or similar captions. In Figure \ref{fig:paraphraser}(a), audio samples $a^i$, $a^j$, and $a^k$ (Three Mel spectrograms at the top), each containing different information, are mapped to the same caption $t$. Similarly, the audio samples $a^x$, $a^y$, and $a^z$ (Three Mel spectrograms at the bottom) are labeled with the caption $v$. This is because, compared to the diverse audio environments that can be recorded, it is difficult to contain detailed and rich text expressions in the annotation process performed by humans. When employing this data for contrastive learning, the same text may be used as negative samples, which can lead to many-to-one mapping issues that adversely impact representation learning performance.

To overcome the many-to-one mapping occurring in the audio-language domain, we propose a novel distance sampling-based paraphraser that uses metrics to compute the distance between two sentences, such as Levenshtein distance or Jaccard similarity, as shown in Figure \ref{fig:paraphraser}(b). The paraphraser that samples a manipulated text $\hat{t}_d$ located at that distance $d$ from a ground-truth $t$ is implemented using ChatGPT, and a few-shot prompt learning scheme is applied to ChatGPT to understand the concept of the quantitative distance $d$.
As a result, as shown in Figure \ref{fig:paraphraser}(c), paraphrased sentences can be obtained for each different audio. 
Because $d_1$ is shorter in ground-truth $t$ than $d_2$, the manipulated text along the $d_1$-surface has fewer variations applied to the original sentence $t$, compared to $d_2$-surface. 
On the other hand, paraphrasing text along the $d_2$-surface can produce results with many variations.

Therefore, the many-to-one problem as shown in Figure \ref{fig:paraphraser}(a) is alleviated and a good representation can be learned. Our method allows us to generate appropriately manipulated text based on the distance metrics, ultimately contributing to more efficient and robust contrastive learning. 

In summary, we propose a novel distance sampling-based paraphraser leveraging ChatGPT to improve audio-language retrieval performance.
The contributions of this study are as follows:

\begin{itemize}
    \item The proposal of a ChatGPT-based paraphraser that can consider quantitative text metrics such as Levenshtein distance or Jaccard similarity.
    \item The proposal of a novel preprocessing and distance-based few-shot prompt sampling method to effectively learn from the distribution of existing text datasets.
\end{itemize}

\section{Related Works}

\subsection{Audio-Language representation learning}

The growing interest in understanding the relationship between audio and text has led researchers to adopt a variety of challenges and research studies.
Since 2020, the Detection and Classification of Acoustic Scenes and Events (DCASE) challenge has featured an audio captioning task, and in 2022, a language-based audio retrieval task was added.

Some studies \cite{Mei2022metric,Koepke2022,xin2023improving,xie2022negative,deshmukh2022audio,laionclap2023} have delved into representation learning using paired audio-text datasets, such as AudioCaps \cite{kim2019audiocaps} and Clotho \cite{drossos2020clotho}. Authors \cite{Mei2022metric,Koepke2022} use audio and text encoders for contrastive learning and demonstrate the efficacy of these encoders in representation learning. Another research direction \cite{xin2023improving} has focused on modifying the loss function to improve performance while also addressing the limitations of current audio-text models. In terms of dataset aspects, Xie \etal \cite{xie2022negative} has demonstrated improved performance by incorporating negative sampling in contrastive learning. However, the lack of available paired datasets still remains challenge. To address this issue, some studies \cite{deshmukh2022audio, laionclap2023} have explored the potential of collecting and refining additional data to improve representation learning, with the aim of expanding the audio-text dataset, but collecting additional data requires additional effort. Additionally, the indirect annotation process involved in generating captions into labels can cause text labels to become similar and reduce quality.

\subsection{Multimodal learning with ChatGPT}

Since the advent of OpenAI's ChatGPT, numerous researchers have been leveraging the capabilities of advanced language models in their studies. Some studies \cite{gilardi2023chatgpt,sun2023chatgpt,moller2023prompt,kuzman2023chatgpt} have explored the potential of ChatGPT for a wide range of natural language processing (NLP) tasks, resulting in notable improvements in the performance of downstream applications such as audio-text retrieval and audio captioning. Furthermore, Gilardi \etal \cite{gilardi2023chatgpt} have established the validity of ChatGPT as an annotator, highlighting its capacity to expedite and simplify the time-consuming process of data generation. Sun \etal \cite{sun2023chatgpt} examined passage re-ranking tasks using instructions such as query generation, relevance generation, and permutation generation. Researchers \cite{moller2023prompt,kuzman2023chatgpt} leveraged the power of ChatGPT to augment data for classification tasks.

\begin{figure*}[t!]
	\centering
	\includegraphics[width=0.95\textwidth]{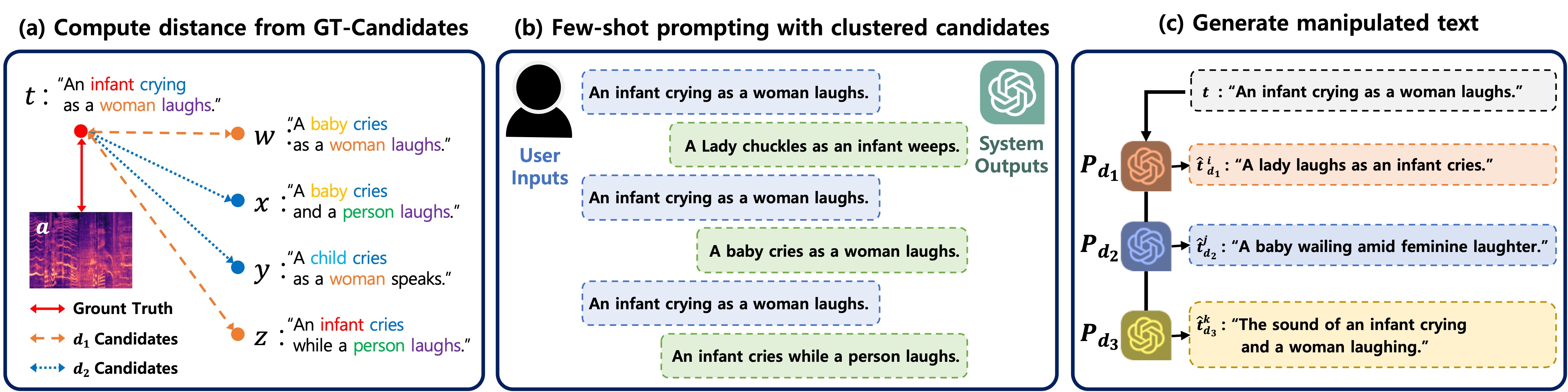}
	\caption{Three-stages pipeline of the proposed Distance Sampling-based paraphraser. (a) $1^{st}$ stage to calcuate a distance between ground truth sentence and candidate sentences, (b) $2^{nd}$ stage to perform few-shot prompting for ChatGPT using the examples clustered by the distance, and (c) $3^{rd}$ stage to generate the manipulated text satisfying the given distance. }
	\label{fig:few-shot} 
\end{figure*}

In the multimodal domain, the use of ChatGPT is gaining momentum. In vision-language domain, researchers \cite{wu2023visual,shen2023hugginggpt} have proposed innovative models capable of executing a variety of tasks by combining ChatGPT with pretrained models. In the audio-language domain, to tackle the limited datasets problem, Mei \etal \cite{mei2023wavcaps} created a new large-scale audio language dataset by using ChatGPT to generate weakly labeled captions and demonstrated excellent performance on related downstream tasks. Studies \cite{wu2023visual,shen2023hugginggpt,mei2023wavcaps} employing ChatGPT in multimodal domain have primarily focused on integrating it with specific tasks. However, there's a noticeable lack of comprehensive analysis on the suitability and efficacy of ChatGPT's outcomes for these tasks. 

Our study explores the influence of a paraphraser leveraging ChatGPT on the performance within a multimodal audio-language retrieval task. Our findings are not just focused on using ChatGPT as the basic form, but extend to provide substantial intuition and insight into the strategic design of optimal calibrations and prompts to maximize efficiency. Therefore, we proposed a novel paraphraser with distance sampling to generate manipulated text samples using ChatGPT, which understands the concept of distance between text samples through prompt learning.

\section{Method}
There are audio and text encoders for audio-text contrastive learning, and both modalities of audio embedding $A$ and text embedding $T$ are optimized through contrastive learning. To differentiate audio-language representation pairs from each other while optimizing audio and text encoders, our distance sampling-based paraphraser $P_d$ generates paraphrased sentences from original text datasets.

\subsection{Audio-Text contrastive learning}

For audio-text contrastive learning, we employ an audio encoder, denoted as $\mathcal{A}$, and a text encoder, denoted as $\mathcal{T}$, to generate the respective embeddings. The audio encoder $\mathcal{A}$ processes the audio sample $a^i$ to yield the audio embedding $A_i$. Similarly, the text encoder $\mathcal{T}$ processes the text sample $t^i$ to produce the text embedding $T_i$, where sample index $i = [1,...,N]$. The encoding process can be formulated as follows:
\begin{gather*}\label{eq:encoding}
A_i = \mathcal{A}(a^i), \quad  T_i = \mathcal{T}(t^i).
\end{gather*}
In a paired audio caption dataset, a pair $(a^i, t^i)$ is considered positive pair and all other combinations $(a^i, t^j)_{j \neq i}$ are considered negative pairs. 
To minimize the distance between positive pairs while maximizing the distance between negative pairs, contrastive learning with NT-Xent loss $\mathcal{L}$  \cite{Mei2022metric, chen2020simple} is performed to generate an optimal audio-text embedding space. The objective function $\mathcal{L}$ is as follows:
\begin{align*}\label{eq:Loss}
\mathcal{L} = -\frac{1}{N} \biggl(& \sum_{i=1}^{N}{\log \frac{\exp(\operatorname{sim}(A_i,T_i)/\tau)}{\sum_{j \neq i}^{N}{\exp(\operatorname{sim}(A_i,T_j)/\tau)} }} \\
+ &\sum_{i=1}^{N}{\log \frac{\exp(\operatorname{sim}(A_i,T_i)/\tau)}{\sum_{j \neq i}^{N}{\exp(\operatorname{sim}(A_j,T_i)/\tau)} } } \biggr),
\end{align*}
where $N$ is the batch size, $i$ and $j$ denote the sample index in a batch, $\tau$ is a temperature hyper-parameter, and cosine similarity is defined as $\operatorname{sim}(A, T)=\frac{A \cdot T}{||A||_2 ||T||_2}$.
Through contrastive learning, a bidirectional audio-text retrieval task can be performed for each audio and text pair.

\subsection{Distance Sampling-based paraphraser}
The details of the distance sampling-based paraphraser using ChatGPT can be described in three-stages, as illustrated in Figure \ref{fig:few-shot}.  

\subsubsection{$1^{st}$ stage - Distance calculation for example clustering} 
First, preprocessing is performed to determine the distance-based distribution, as shown in Figure \ref{fig:few-shot}(a). Let us assume that there is a ground truth sentence $t$ for audio $a$, and that there is a set of candidate sentences  $w, x, y, z$ that have been paraphrased from the sentence $t$. All sentences are morphologically analyzed to extract only nouns $N$ and verbs $V$. The extracted nouns $N$ and verbs $V$ sets are then converted into union sets $U_* = N_* \cup V_*$ where $ * = \{t, w, x, y, z\}$. The distance $d(U_t,U_{* \setminus t})$ between the ground truth $t$ and each candidate $({* \setminus t})$ are calculated. Finally, candidates are clustered according to the distance $d$. In this paper, the distance function $d$ is the inverse of the Jaccard similarity \cite{jaccard1912distribution}.

\subsubsection{$2^{nd}$ stage - Few-shot prompting examples of ChatGPT} 

ChatGPT is effective at providing answers to new input based on the few-shot examples. As shown in Figure \ref{fig:few-shot}(b), to perform a few-shot prompting for ChatGPT, candidate text is sampled from clusters of similar distances $d$ and provided as system output, and user input is provided as ground truth. 

\subsubsection{$3^{rd}$ stage - Text manipulation with distance constraints} 
With few-shot prompting, the ChatGPT-based paraphraser $P_d$ performs manipulations on new sentences. At this point, the paraphrased sentence undergoes a change similar to the distance $d$ of the few-shot examples input as $2^{nd}$ stage. In Figure \ref{fig:few-shot}(c), when ``\textit{An infant crying as a woman laughs.}'' is ground-truth sentence $t$, ``\textit{A lady laughs as an infant cries}'',  ``\textit{A baby wailing amid feminine laughter.}'', and ``\textit{The sound of an infant crying and a woman laughing.}'' are samples manipulated with the ChatGPT-based paraphraser $P_d$ with the distance constraints $d_1 < d_2 < d_3$, respectively.

\section{Experiments}
\subsection{Experimental setup}
As the audio-language dataset, Audiocaps \cite{kim2019audiocaps} is used. 
The dataset is partitioned into distinct train, validation, and test sets. Specifically, the training dataset contains a single caption for each audio sample, whereas the validation and test sets have five corresponding captions per audio sample. Since our proposed method requires paraphrased candidates for ground truth sentences, as depicted in Figure \ref{fig:few-shot}(a), the validation dataset is used to construct text cluster according to distance $d$. For the audio-text representation learning model, the baseline model \cite{Mei2022metric} is used. 
The baseline model is applied in a two phase training scheme consisting of (1) pretraining phase and (2) finetuning phase. In the pretraining phase, the model is trained using both the original dataset and the manipulated dataset with distance constraints applied. After the pretraining phase, the model is finetuned using only original dataset.
The model is pretrained using the AdamW \cite{loshchilov2017decoupled} optimizer with 50 epochs with a batch size of 32 and a learning rate of $1 \times 10^{-5}$, and finetuned by an additional 20 epochs with a learning rate of $1 \times 10^{-6}$.

\begin{table}[t]
\centering
  \caption{Results of audio-text retrieval on AudioCaps. Proposed ($d$, $n$) means that $d$ is normalized distance and $n$ is the number of few-shot samples. R@$k$ (recall at rank $k$) in bold and underline is the highest and the next highest scores.} 
  \begin{adjustbox}{width=0.45\textwidth}
  \label{tab:result}
  \begin{tabular}{c ccc ccc}
    \hline   
      \hline

    \multirow{2}{1em}{Model}& \multicolumn{3}{c} {(a) Text-to-Audio} & \multicolumn{3}{c} {(b) Audio-to-Text}\\
    \cmidrule(lr){2-4} \cmidrule(lr){5-7}
     & R@1 $\uparrow$ & R@5 $\uparrow$ & R@10 $\uparrow$ & R@1 $\uparrow$ & R@5 $\uparrow$ & R@10 $\uparrow$ \\
    \hline
    Baseline \cite{Mei2022metric} 
                        & 34.78             & 70.78	            & 83.76	            & 40.78	            & 76.11             & 86.22 \\
     \hline

    Text augmentation \cite{wei2019eda} 
                        & 36.18             & 70.38             & 83.56             & 42.33             & 74.89	            & 86.00	\\

   Audio Captioning \cite{mei2021audio} 
                        & \underline{36.27} & 70.49             & \underline{84.04} & 41.78             & 75.56             & \underline{87.11} \\
     
   Wavcaps \cite{mei2023wavcaps} 
                        & \bf{36.87}        & \underline{71.42} & \underline{83.87} & 44.00             & \underline{75.56} & 86.67\\
     \hline

    Proposed (10\%, 10) & 35.44             & \bf{71.44}        & \bf{84.22}        & 42.89             & 75.89             & \underline{87.11} \\
    Proposed (10\%, 30) & 36.24	            & 71.24	            & 83.73             & \bf{44.67}        & \bf{77.11}        & \bf{87.22}	 \\
    Proposed (10\%, 50) & 36.16	            & 70.89	            & 83.80             & \underline{44.22} & \underline{76.22} & 86.78	 \\
    \hline
    \hline

\end{tabular}
\end{adjustbox}
\vspace{-0.2cm}
\end{table}

\subsection{Model performance}
Comparative experiments with other data augmentation models were performed and the result are summarized in Table \ref{tab:result}. The baseline \cite{Mei2022metric} is a model trained using only Audiocaps dataset. The text augmentation \cite{wei2019eda} uses data augmented with EDA, which is commonly used in NLP. The audio captioning model \cite{mei2021audio} uses audio captions generated through a pretrained audio captioning model as additional data. The Wavcaps-based method \cite{mei2023wavcaps} augments existing captions using a method of generating descriptions with ChatGPT. 

As shown in Table \ref{tab:result}, the proposed method showed good performance, achieving highest or second scores in almost recall at rank $k$ (R@$k$) compared to other methods. Specifically, in Text-to-Audio experiment, using the augmented dataset from the proposed paraphraser constrained with $d=10\%$ and $n=10$ outperforms the others, but the dataset with $d=10\%$ and $n=30$ are most efficient in Audio-to-Text scenario.

To verify the distance mechanism of the proposed paraphraser, 
Table \ref{tab:distance} shows Jaccard similarity calculated for texts manipulated by paraphraser with different distance constrains applied. The larger the distance constrains, the smaller the Jaccard similarity.

\begin{table}[t]
\centering
  \caption{Jaccard similarity with respect to normalized distance $d$ and the number of few-shot samples $n$.}
  \begin{adjustbox}{width=0.4\textwidth}
  \label{tab:distance}
  \begin{tabular}{c ccc}
    \hline   
      \hline
    Jaccard similarity  & \multirow{2}{3em}{$n = 10$} & \multirow{2}{3em}{$n = 30$} & \multirow{2}{3em}{$n = 50$} \\
    (mean $\pm$ std)      & & & \\
    \hline
    $d = 10\%$ & 0.66 $\pm$ 0.18 & 0.70 $\pm$ 0.19 & 0.69 $\pm$ 0.19 \\
    $d = 50\%$ & 0.38 $\pm$ 0.18 & 0.43 $\pm$ 0.19 & 0.45 $\pm$ 0.19 \\
    $d = 90\%$ & 0.24 $\pm$ 0.14 & 0.23 $\pm$ 0.14 & 0.25 $\pm$ 0.14 \\
     \hline
    \hline
\end{tabular}
\end{adjustbox}
\vspace{-0.3cm}
\end{table}

\subsection{Ablation study}
Table \ref{tab:abalation_test} shows an ablation study to investigate the effects of the normalized distance $d$ and the number of few-shot examples $n$ on the performance of the proposed distance sampling-based paraphraser. Specifically, we selected 10, 30, and 50 samples to perform few-shot prompt learning for the ChatGPT, and experimented with normalized distances of 10\%, 50\%, and 90\%. The proposed paraphraser achieved the best performance with few-shot prompting using 30 samples with 10\% distance. 

From the perspective of the number of few-shot samples, 10 samples are not enough to perform the few-shot prompting, and 50 samples is too many and noisy, making it difficult to build a concentrated data distribution with a small standard deviation. However, 30 samples are sufficient to perform few-shot prompting, and the proposed paraphraser produces manipulated sentences that match the distribution of the original dataset.

Additionally, using few-shot samples over a short distance, the paraphraser acts as an interpolator to create a fine grid in intra distribution, but if the paraphraser is constrainted to a long distance, it becomes an extrapolator that samples between independent samples in inter distribution.
The characteristics of the paraphraser according to the distance constraints allows us to effectively manipulate text samples by controlling their original distribution.
These findings provide important insights for researchers and practitioners seeking to improve the performance of few-shot prompting.

\begin{table}
  \caption{Results from the number of few-shot samples and distance for the proposed model. The best scores are \textbf{bold}.}
  \begin{adjustbox}{width=0.45\textwidth}
  \label{tab:abalation_test}
  \begin{tabular}{cc ccc ccc}
    \hline
        \hline   
    \multirow{2}{5em}{Normalized distance $d$} & \multirow{2}{6em}{\# of few-shot samples $n$} & \multicolumn{3}{c} {(a) Text-to-Audio} & \multicolumn{3}{c} {(b) Audio-to-Text }\\
    \cmidrule(lr){3-5} \cmidrule(lr){6-8}
    & & R@1 $\uparrow$ & R@5 $\uparrow$ & R@10 $\uparrow$ & R@1 $\uparrow$ & R@5 $\uparrow$ & R@10 $\uparrow$ \\
\hline

    \multirow{3}{2em}{10\%} & 10 & 35.44 & \bf{71.44} & \bf{84.22} & 42.89 & 75.89 & 87.11 \\
                            & 30 & \bf{36.24} & 71.24 & 83.73 & \bf{44.67} & \bf{77.11} & \bf{87.22} \\
                            & 50 & 36.16 & 70.89 & 83.80 & 44.22 &  76.22 & 86.78 \\
                          
     \hline   
    \multirow{3}{2em}{50\%} & 10 & 35.07 &	71.20 &	84.04 &	44.56 &	76.00 & 85.78 \\
                            & 30 & 35.67 & 70.47 & 83.36 & 44.00 & 75.33 & 85.11 \\
                            & 50 & 35.82 & 71.16 & 83.96 & 42.22 & 74.89 & 87.00 \\
     \hline   
    \multirow{3}{2em}{90\%} & 10 & 36.07 & 70.40 & 83.62 & 43.56 & 74.00 & 86.00 \\
                            & 30 & 35.67 & 70.62 & 83.56 & 42.67 & 74.89 & 85.44 \\
                            & 50 & 36.07 & 70.67 & 83.53 & 44.22 & 75.22 & 85.78 \\
     \hline   
  \hline
\end{tabular}
\end{adjustbox}
\vspace{-0.5cm}
\end{table}

\section{Conclusion}
In this study, we propose a novel distance sampling-based paraphraser designed to effectively generate manipulated sentences for audio-text retrieval tasks. By employing a few-shot learning approach and utilizing distance-contrainted sampling, our model is able to generate paraphrased sentences with varying degrees of transformation based on the chosen distance. Our experimental results demonstrate the superior performance of our proposed model in comparison to other models within the same domain. Furthermore, Our ablation studies show that few-shot prompting of a few or many samples biases the data distribution away from the original distribution. In addition, the proposed paraphrasers can also act as different functions, such as interpolator or extrapolator, depending on the distance of text clusters. 

\begin{acks}
This work was supported by Institute of Information \& communications Technology
Planning \& Evaluation (IITP) grant funded by the Korea government(MSIT) (No. 2022-0-00320, Artificial intelligence research about cross-modal dialogue modeling for one-on-one multi-modal interactions)
\end{acks}

\bibliographystyle{ACM-Reference-Format}
\balance
\bibliography{sample-base}

\end{document}